# An Open Research Dataset of the 1932 Cairo Congress of Arab Music

Baris Bozkurt

College of Interdisciplinary Studies, Zayed University, Dubai, United Arab Emirates

## 1. Introduction

This paper presents an open research dataset (ORD-CC32) designed to support computational (ethno)musicological studies of Arab music. The dataset includes metadata, melodic and rhythmic mode tags, tonic annotations, and intonation/melody features extracted from historical recordings of the 1932 Cairo Congress of Arab Music, a landmark event in the history of Arab music. The recordings originate from a diverse range of Arab regions and were released as an 18-CD box set (sound restored by Luc Verrier), accompanied by a trilingual booklet in French, English, and Arabic (Lambert & Cordéreix, Eds., 2015). This booklet provides detailed information about the congress, the musicians, recording conditions, and various musical features (such as melodic and rhythmic modes (although incomplete)), which served as the foundation for the metadata in our dataset. Despite the historical and cultural importance of this collection, there is a surprising lack of machine-readable resources and computational studies dedicated to it. Our work addresses this gap by sharing a dataset composed of metadata and features computed from the recordings. In the following sections, we first discuss the congress's significance and the role of this research dataset.

### 1.1 The 1932 Cairo Congress of Arab Music and its significance

The 1932 Cairo Congress of Arab Music, was a seminal event that brought together musicians, scholars, and cultural delegates from across the Arab world and Europe to discuss, document, preserve, and systematize Arab music traditions. Held under the patronage of King Fuad I, the congress was both a scientific and political undertaking. At a time when many Arab nations were emerging from Ottoman influence, the congress functioned as a symbolic and strategic act of cultural diplomacy. It projected a unified Arab identity rooted in a shared musical heritage, asserting cultural sovereignty and intellectual agency on the global stage. The inclusion of international delegates (from Turkey, Persia, and Western Europe) further reflects the congress's embeddedness in global political and cultural currents. The event symbolized a moment aiming to position Arab music within global academic and artistic discourses (Bensignor, 2015). It featured performances, debates, and comparative studies of musical systems.

"The Cairo Congress of Arab Music 1932" is the earliest international event that brought together representative music groups from various Arab music traditions. With the aim of "reviving and systematizing Arab music", numerous musical performances were recorded during the conference which were recently subject to sound restoration and were published (for the first time as a whole) together with a book in English, French, and Arabic (Original text by B. Moussali; J. Lambert & P. Cordéreix, Eds., 2015). The recording collection comprises 333 tracks classified by country (Egypt, Iraq, Syria, Turkey, Algeria, Morocco, Tunisia). The metadata of the tracks are available in printed form (in the book) and includes the titles, instrumentation, musical modes, musical forms, etc. While the coverage of metadata is not complete, music mode tags of many of the recordings are available which is indispensable for computational studies on tuning/intonation and melodic patterns.

The contribution of the congress on cultural heritage preservation has been significant. To define a unified theory for Arab music (and "establishing a fixed musical scale"), a dedicated committee has been set up. The committee could not reach a final commonly agreed proposal (for a unified Arab music theory). Interestingly, the organizers were aware that computational analysis of the recordings would be of utmost use for a well-grounded study for tuning and temperament and the technological deficiencies of the time which could be overcome in the future: "…The 171 discs … often include tuning instruments applied and the A frequency (435 Hz at 20C). This disposition is made to assist future researchers on musical scales…"

(Moussali, 2015, page 112). The technological deficiencies (for computing pitch reliably) are indeed almost over now. Yet, we still lack the computational studies provisioned by the organizers in 1932.

**1.2 The role of research corpora for recordings of The 1932 Cairo Congress of Arab Music**

Research corpora are fundamental for the computational study of music, serving as the backbone for empirical analysis, algorithmic development, and theoretical inquiry. The creation of such corpora is not merely a mechanical task but a research endeavor in itself, requiring careful consideration of the musical, cultural, and historical specificities of the tradition being studied. For Arab music, this involves accounting for its unique melodic and rhythmic modes (maqamat and iqa'at), microtonal nuances, and the oral transmission practices that have shaped its evolution. The dataset presented here is tailored to the culture specific aspects, providing structured, machine-readable resources that align with the intrinsic characteristics of Arab music while facilitating cross-cultural computational studies.

Computational analysis has the potential to make implicit knowledge from musicians explicit, uncovering patterns and practices that may not be formally documented. For example, the microtonal inflections and improvisational techniques inherent in Arab music often reside in the tacit knowledge of performers. By leveraging computational tools, a research dataset has the potential to help codify such knowledge, offering insights into performance practices, intonation systems, and regional variations. This not only enriches academic understanding but also supports pedagogical applications, ensuring that this cultural heritage is preserved and transmitted accurately.

A primary aim of this dataset is to push forward the state of the art in computational musicology and Music Information Retrieval (MIR) for this music tradition. By offering high-quality acoustic features, metadata, and annotations, the dataset will enable researchers to develop and test algorithms for tasks such as automatic mode recognition, tonic identification, and tuning analysis. These advancements are not only technically significant but also culturally meaningful, as they allow for the systematic exploration of Arab music traditions. The dataset bridges a critical gap in MIR, where non-Western music traditions remain underrepresented, and provides a foundation for future research in computational (ethno)musicology.

The dataset holds significant promise as a foundational resource for scholars across a wide array of disciplines, not only in musicology and computational music analysis but also in history, cultural studies, postcolonial studies, and political science.

Furthermore, by openly sharing this dataset, we aim to prevent duplication of efforts in building corpora, a concern for research institutes with limited resources globally. Next, we summarize the current state of open research corpora for Arab music and studies carried on these corpora.

## 2. Related Work and Research Gap

**2.1 Open research datasets for Arab Music**

The field of computational musicology, especially concerning under-represented traditions like Arab music, has historically confronted a significant challenge: the scarcity and fragmented nature of accessible research corpora. A comprehensive review of freely available corpora in the Arabic language (including text, audio or image datasets) has been carried by Ahmed et al, 2022. Among 63 corpora reviewed, we only find one open-music dataset by Chaachoo et al, 2018. The review concludes that the existing resources for Arabic corpora are often scattered and not maintained.

Focusing on open music datasets for Arab music, by the date of this publication, several music research datasets are available. The *Arab-Andalusian Music Corpus* (Sordo et al., 2014) offers 338 recordings from Morocco, annotated with cultural-specific tags like *nawba* and *mizán*, alongside lyrics. Its successor (Chaachoo et al., 2018) expands this effort with 164 concert recordings, MusicXML scores, and pitch annotations. Complementing these, the *Arab-Andalusian Music Lyrics Dataset* (Sordo et al., 2018) provides lyrics in Arabic and transliteration, enabling studies of text-melody relationships. The *Arabic Poetry and*

*Melody Dataset* (Shahriar & Al Roken, 2022) pairs poems labeled by emotion with melodies in various maqams, while the *Arabic Music Genre Classification Dataset* (Almazaydeh et al., 2022) offers 1,266 genre-labeled clips for machine learning tasks. Lastly, the *Parallel Corpus for Arab Music Improvisation* (Al-Ghawanmeh et al., 2023) is designed for generative AI applications, including vocal-instrumental improvisation data. The following table presents a summary of these resources.

| Name of the Dataset | Authors/Citation | Content | Resource Link |
|---|---|---|---|
| **Arab-Andalusian Music Corpus** | Sordo et al. (2014) | 338 audio recordings with metadata, lyrics, and music scores; nawba, tab', mizán annotations. | http://dunya.compmusic.upf.edu/andalusian/ |
| **Arab-Andalusian Music Corpus** | Chaachoo et al. (2018) Repetto et al. (2018) | 164 concert recordings (125+ hours), MusicXML scores, pitch data, JSON metadata. | https://zenodo.org/records/1291776 |
| **Arab-Andalusian Music Lyrics Dataset** | Sordo et al. (2018) | Lyrics in Arabic script and transliteration, aligned with audio via MusicBrainz IDs (TSV/JSON). | https://zenodo.org/records/3337623 |
| **Arabic Poetry and Melody Dataset** | Shahriar & Al Roken (2022) | Arabic poems labeled by emotion paired with maqam-based melodies (e.g., Saba, Seka, Ajam). | https://github.com/sakibsh/Arabic-Poetry-Melody |
| **Arabic Music Genre Classification Dataset** | Almazaydeh et al. (2022) | 1,266 audio clips (30s each) across 5 genres (WAV format). | https://www.kaggle.com/datasets/araraltawil/audio-classifier |
| **Parallel Corpus for Arab Music Improvisation** | Al-Ghawanmeh et al. (2023) | 6,991 vocal-to-instrumental sentences + 717 instrumental improvisations for machine translation. | https://github.com/FadiGhawanmeh/AMICOR |

**Table 1:** Open Arab music datasets

These datasets excel in cultural specificity, particularly the Arab-Andalusian corpora. Their open-access availability (via Zenodo, GitHub, and Kaggle) fosters computational research for Arab music. For example, the Arab-Andalusian music corpus (Chaachoo et al, 2018) has served as the essential data foundation for initiating computational studies on this under-represented music tradition, enabling both MIR tasks like automatic *nawba* recognition (Pretto, et al, 2018) and musicological explorations like the analysis of melodic patterns related to centonization theory (Nuttall, et al, 2023).

**2.2 Dataset gap for cross-regional/cultural and historical recordings**

Despite a growing number of open datasets for Arab music, there remains a notable gap in cross-regional and historical computational studies. Most existing corpora focus on individual traditions (such as Arab-Andalusian or Egyptian music). Their potential for use for comparative analysis across the diverse musical practices of the Arab world is very limited. Consequently, there is a lack of computational research that systematically explores regional variations in Arab music traditions, particularly in relation to intonation and tuning practices.

The dataset introduced in this paper addresses this gap by focusing on recordings from the 1932 Cairo Congress that includes representative recordings from music traditions of Algeria, Egypt, Iraq, Morocco, Syria, Tunisia and Turkey. We argue that computational analyses of this dataset to characterize regional differences would be highly aligned with the heterogeneity intrinsic to Arab musical practice and could provide deeper insights into the phenomenon of musical scales, or maqamat, as they exist in performance.

To date, computational studies of the Cairo Congress recordings are scarce. Historical scholarship has examined the congress's role in the standardization of Arab music theory and its cultural implications (e.g., Bois, 1990; Bensignor, 2015). The pitch content and intonation practices embedded in these recordings remain largely unanalyzed using contemporary computational tools. Our dataset enables such analysis, offering researchers structured, machine-readable access to audio features, tonic annotations, and mode tags.

Furthermore, this dataset presents an opportunity to compare historical performance practices with contemporary ones, thereby enabling diachronic studies of tuning, modality, and ornamentation. It opens pathways for investigating how musical scales have evolved over time and across regions.

A useful point of reference for such work is Elsner (2024), who examines the complexity and regional branches of the maqam in Azerbaijani mugham music. Elsner highlights how creative performance practices and regional variation challenge the notion of a singular, unified system. Similarly, our dataset can support investigations that unpack how regional traditions, performer agency, and oral transmission have contributed to the diversity of Arab tuning systems. Rather than imposing a generalized model, the dataset facilitates empirical approaches to understanding how variation itself is a fundamental characteristic of Arab music.

In the next section, we provide a brief presentation of the dataset content.

## 3. Dataset Content

We share the open research dataset of 1932 Cairo Congress recordings, shortly named as **ORD-CC32**, on Zenodo: https://zenodo.org/records/15682346.

The dataset contains the following features, metadata, tags and plots:

- MusicBrainz ID of the track which makes accessible to the following metadata through musicBrainz API: Title, Length, Track artist, Release title, Release artist, Release group type, Country/Date, Label, Catalog number. The main release record on musicBrainz is accessible through: https://musicbrainz.org/release/25d66aec-1463-4a0b-a331-d3bbe587e71c
- Pitch series and pitch confidence extracted using pYIN[1] (Mauch & Dixon, 2014), Crepe (Kim et al, 2018) and predominant melody makam (Atli et al, 2014) (analysis time step (for all extractors): 10 milliseconds).
- Manually labeled tonic segment time location in the recording and the frequency estimated (for a subset (64 out of 333) of recordings)
- Pitch histograms (three versions: regular, octave warped, tonic aligned octave warped) computed using the three methods and a mean histogram averaging all histograms using different pitch extractors. Pitch histograms are both stored as series/arrays and also provided as plots to help musicologists perform visual inspection without the need of running any Python code.
- Automatically extracted histogram maxima locations (from the mean histogram) and pitch scale intervals

---

[1] Implementation in the Librosa library have been used for extraction: https://librosa.org/doc/0.11.0/generated/librosa.pyin.html

- Maqam tag, rhythmic mode tag, musical form tag together with tags specifying if the recording is vocal-only, instrument-only or a mixture.

No audio recording is included in the dataset due to property rights restrictions. The dataset also contains a folder that includes the Python code to extract all pitch features specified above, to retrieve comprehensive metadata including artists, releases, and recording details using MusicBrainz MBIDs. This facilitates updates when new tags are added to the dataset or re-creating all acoustic features from audio recording.

Table 2 provides the number of tracks available for a subset of the region-maqam pairs.

| region | maqam | number of tracks | region | maqam | number of tracks | region | maqam | number of tracks |
|---|---|---|---|---|---|---|---|---|
| Algeria | raml al-ashiyya | 11 | Egypt | not available | 54 | Iraq | not available | 6 |
| Algeria | raml al-maya | 5 | Egypt | bayati | 27 | Iraq | ibrahimi | 6 |
| Algeria | rasd al-dhil | 2 | Egypt | rast | 15 | Iraq | mansuri | 6 |
| Algeria | ghrib | 1 | Egypt | hijaz | 11 | Iraq | mukhalif | 6 |
| Algeria | husayni | 1 | Egypt | saba | 11 | Iraq | bherzawi | 4 |
| Algeria | iraq | 1 | Egypt | nahawand | 10 | Iraq | dukah | 4 |
| Algeria | jaharkah | 1 | Egypt | sikah | 4 | Iraq | bayati | 3 |
| Algeria | maya | 1 | Egypt | sikah arabi | 4 | Iraq | husayni | 3 |
| Algeria | mazmum | 1 | Egypt | ushshaq misri | 4 | Iraq | lami | 3 |
| Algeria | mujannaba | 1 | Egypt | zunjuran | 4 | Iraq | rast | 3 |
| Algeria | rast | 1 | Egypt | iraq | 3 | Iraq | humayun | 2 |
| Algeria | sikah | 1 | Egypt | ajam ushayran | 2 | Iraq | saba | 2 |
| Algeria | zidan | 1 | Egypt | awj | 2 | Iraq | sikah | 2 |
| Morocco | hijaz al-kabir | 1 | Egypt | isfahan | 2 | Iraq | hakimi/huzam | 1 |

**Table 2:** Number of tracks for region-maqam pairs (this table is incomplete due to space considerations, the complete table available as a csv file among the dataset files on Zenodo)

|  | Algeria | Egypt | Iraq | Morocco | Syria | Tunisia | Turkey |
|---|---|---|---|---|---|---|---|
| # of tracks | 28 | 168 | 59 | 30 | 2 | 40 | 6 |

Table 3: Number of recordings/tracks per region

One particular problem is the metadata on musicBrainz being in French rather than Arabic. This is a downside considering culture specificity which may be directly corrected/added on musicBrainz. Once the content in Arabic is available on musicBrainz, updates could be easily performed via running a script in the Python code base shared with this dataset.

| path | musicBrainz link | Track Title | Tags | vocal-only | tonic_Hz | artist | region |
|---|---|---|---|---|---|---|---|
| CD 1/01. | https://musicbrainz.org/recording/d64461bb-c41b-4396-a671-caf846205b34 | Les nuits d'amour / Ô mon Commensal | ['mode: hijaz', 'rhythm: muhajjar misri (14/4)', 'rhythm: murabba misri (13/4)', 'tonic_secs: 0.9-1.6'] | FALSE | 135.22 | Darwîsh Muhammad al-Harîrî | Egypt |

Table 4: Sample data entry (a complete table for all recordings provided in the dataset )

**Les nuits d'amour / Ô mon Commensal**
~ Recording by Darwîsh Muhammad al-Harîrî

Overview | Fingerprints | Aliases | Tags | Reviews | Details | Edit

**Appears on releases**

| # | Title | Length | Track artist | Release title | Release artist | Release group type | Country/Date | Label | Catalog# |
|---|---|---|---|---|---|---|---|---|---|
| Official | | | | | | | | | |
| 1.1 | Les nuits d'amour / Ô mon Commensal | 3:50 | Darwîsh Muhammad al-Harîrî | Congrès de musique Arabe du Caire 1932 | Various Artists | Album + Compilation | FR 2015-03 | BnF Collection (Bibliothèque nationale de France) | BNF 01 CD-01-18 |

Overview | Fingerprints | Aliases | **Tags** | Reviews | Details | Edit

**Genres**
There are no genres to show.

**Other tags**

| mode: hijaz | + − | 1 |
| rhythm: muhajjar misri (14/4) | + − | 1 |
| rhythm: murabba misri (13/4) | + − | 1 |
| tonic_secs: 0.9-1.6 | + − | 1 |

Figure 1: MusicBrainz view for the sample in Table 4 (https://musicbrainz.org/recording/d64461bb-c41b-4396-a671-caf846205b34)

Unique maqam labels in the dataset are: 'hijaz', 'hijaz al-kabir', 'hijaz divan', 'bayati', 'saba', 'rast', 'ibrahimi', 'sikah/segah', 'sikah arabi', 'lami', 'mansuri', 'mukhalif', 'awj', 'ushshaq', 'banjgah', 'bherzawi', 'humayun', 'dukah', 'husayni', 'hakimi/huzam', 'khan abat', 'tahir', 'hadidi', 'hulaylawi', 'jabburi', 'awshar', 'rahat al-arwah', 'hicazkar', 'kurdili hicazkar', 'hicazkar', 'raml al-ashiyya', 'raml al-maya', 'ghrib', 'maya', 'rasd al-dhil', 'zidan',

'mujannaba', 'mazmum', 'jaharkah', 'iraq', 'istihlal', 'dhil', 'isbaayn', 'shahnaz', 'rahawi', 'nawa', 'muhayyar sikah', 'muhayyar iraq', 'isfahan', 'nahawand', 'shawrak', 'nayriz', 'kurdan', 'ushshaq misri', 'zunjuran', 'ajam', 'ajam ushayran', 'udhdhal', 'farahnak'.

Unique rhythmic mode labels in the dataset are: 'samai', 'samai darij (3/4)', 'msaddar (6/4)', 'mukhammas misri (16/4)', 'abyat (4/4)', 'dawr kabir mawlawi (14/2)', 'masmudi (8/4)', 'basit (12/8)', 'zaffe (8/4)', 'btayhi', 'wahda baghdadiyya (4/4)', 'samai sarband (3/8)', 'awfar mawlawi (9/4)', 'zeybekli (9/8)', 'quddam (6/4)', 'quddam (6/8)', 'wahda (4/4)', 'ay nawasi', 'samai thaqil (10/8)', 'qasid (4/4)', 'inqlab (5/4)', 'basit (6/4)', 'msaddar (3/4)', 'barwal (2/4)', 'awfar misri (19/4)', 'samah (36/8)', 'btayhi (8/8)', 'duyek (8/4)', 'agir duyek (8/4)', 'masmudi (4/4)', '(4/4)', 'qaim wa-nusf (8/8)', 'btayhi (4/8)', 'barwal', 'sittata ashara misri (32/4)', 'nawakht', 'nawakht hindi (16/4)', 'masmudi (8/8)', 'dawr kabir mawlawi (14/4)', 'barwal (4/4)', 'sufyan (7/8)', 'qaim wa-nusf (4/4)', 'nawakht misri (7/8)', 'ardawi', 'yuruk samai (6/8)', '(6/8)', 'msaddar', 'yugrug (12/8)', 'qaim wa-nisf (4/4)', 'darj (6/8)', 'tushiya (4/4)', '(8/8)', 'btayhi (4/4)', 'muhajjar misri (14/4)', 'darj', 'murabba misri (13/4)', '(8/4)', 'msaddar (4/8)', 'sadawi (6/8)', 'tarabulusi', 'insiraf (5/8)', 'quddam (9/4)', 'quddam (3/4)', 'msaddar (4/4)', 'khlas (6/8)', 'insiraf', 'zari (8/8)', 'nawakht misri (7/4)', 'mdawwar hawzi', 'insiraf (5/4)', '(6/4)', 'shanbar misri (48/4)', 'kursi (5/4)', 'qaim wa-nisf (8/4)', 'sengin samai', 'shanbar misri (14/4)', 'dawr rawan mawlawi (14/8)', '(2/4)'.

Unique musical form labels are: 'taqsim', 'qasida', 'mawwal', 'mandra'.

## 4. Case Study: Comparative Maqam Tuning Analysis

In this section, we present a limited demonstration of a case study to exemplify potential uses of the dataset.

### 4.1 Tuning and temperament in Arab music

Arab music is recognized for its distinctive soundscapes and its microtonal nature. The use of intervals smaller than a semitone, known as microtones or quarter tones is a fundamental difference significantly contributing to the unique and expressive qualities of Arab music melodies. These fine pitch variations enable a richness and complexity that is central to the aesthetic appeal of Arab music. The maqam system is far more than a simple scale; it is a comprehensive musical mode that dictates melodic structures, improvisation, and performance practice. Each maqam includes specific scale tunings, a distinctive melodic vocabulary, hierarchy of pitches and their specific use in setting the form of a musical piece and expected patterns of modulation. We will however limit our discussions to only the scale components to focus on a single specific aspect of maqam music that could be studied using the dataset presented here.

The microtonal structure, tuning, and temperament of Arab music have been subjects of theoretical debate for centuries, with discussions becoming particularly prominent during the 19th and 20th centuries. It is not the aim of this paper/section to provide a comprehensive review of these theoretical debates. We provide a very short summary with pointers for the interested reader and set the base for rationale for the need for computational studies of tuning and temperament.

The current Arab tone system traces its historical roots to the work of Farabi (d. 950 CE), an influential scholar who proposed a 25-tone unequal scale. Safi al-Din al-Urmawi developed, in the thirteenth century, a 17-tone scale (again unequal division) that served as a universally accepted theoretical basis for music throughout the Islamic world for many centuries. In the 19th century, key figures such as Muḥammad al-ʿAṭṭār and his student Mīkhāʾīl Mušāqa played pivotal roles in the emergence of what is now called "modern" Arabic music theory, especially regarding the establishment of the 24-tone equal temperament system. Their work marked a shift towards a standardized approach to tuning in Arab music, influencing subsequent theoretical and practical developments (Maraqa, 2023). A significant factor in the adoption and theoretical prominence of 24-TET was the growing influence of Western musical thought in the Middle East during the 19th and early 20th centuries. This system was notably discussed at the 1932 Cairo Congress of Arab Music, where it faced both support and opposition. Turkish musicologist Rauf Yekta, for example, rejected the 24-tone equal temperament, advocating instead for the preservation of traditional makam systems and highlighting the tension between Westernization and local musical traditions (Öğüt, 2024). A significant consensus among ethnomusicologists and theorists views the 24-tone scale primarily as a theoretical construct or a "conceptual map" that musicians use to discuss and compare intervals, rather than a precise reflection of what is performed.

Arab music tuning and temperament studies reveal a dynamic interplay between tradition and modernity, with ongoing debates about the adoption of equal temperament, the preservation of modal systems, and the impact of tuning on musical perception and identity. These discussions are informed by both historical scholarship and contemporary analytical methods, reflecting the rich and evolving nature of Arab music theory and practice. For example, Ghrab, 2005, examines how Western scholars have studied musical intervals in "Arabic music" over the past three centuries tracing the historical evolution of this Western interest, connecting it to the broader relationship between European and Arab cultures. The study states that Western perceptions of "Arabic music" intervals have evolved, often influenced by Western theoretical frameworks and the changing relationship between European and Arab cultures.

Practical attempts to reconcile theory and performance, such as Julien Jalâl Ed-Dine Weiss's invention of a qānūn in just intonation, highlight ongoing efforts to bridge the gap between theoretical models and actual musical practice (Pohlit, 2017).

Shehadeh, et al, 2003 discusses the extensive regional variants in the maqam system across the Arabic world and beyond, emphasizing that there is not a single unified maqam system but rather diverse regional frameworks for tuning and temperament. Touma, 1971, Explores the maqam phenomenon across North Africa, the Near East, and Central Asia, noting the existence of three principal spheres (Turkish, Persian, Arabian) and the regional diversity in the organization and performance of maqam/makam music. The dataset we present with this paper provides a unique resource to study such regional variations.

**4.2 Pitch histograms for computational study of tuning and temperament**

Pitch histograms[2] and/or distributions are powerful tools for empirically analyzing tuning and temperament across diverse musical traditions, supporting both theoretical comparison and practical applications in music information retrieval and instrument tuning.

One of the early uses of pitch histograms in MIR is in Tzanetakis, et al 2003, where the authors evaluate the use of pitch histograms in the context of automatic musical genre classification and conclude that they can be effectively used for this task. Tzanetakis, et al., 2007 considers pitch histograms as essential computational tools in ethnomusicology, especially valuable for empirical investigation of tuning systems in various musical traditions. The paper emphasizes that pitch histograms enable researchers to objectively analyze theories of tuning and interval structures, particularly in music cultures utilizing intervals smaller than Western semitones, such as Turkish makam and Indian classical music. By generating these histograms from audio recordings, computational methods help distinguish systematic tunings from variations like vibrato or out-of-tune performance, thus providing robust empirical evidence to explore and validate ethnomusicological hypotheses and theories.

A large number of studies have considered use of pitch histograms for tuning analysis for a specific music culture. Some examples are summarized below.

Bozkurt 2008 has proposed the use of aligned mean pitch histograms that are obtained by averaging histograms that are automatically aligned (to maximize match in the distribution space) to serve as basis for a study of pitch scales. Bozkurt et al 2009, extended this work via using the aligned mean histograms of select recordings of Turkish Makam music for the comparison of measured relative pitches from performances with the tones of various proposed theoretical maqam scales, helping to evaluate how well these theories represent actual musical practice.

---

[2] A pitch histogram is a graphical representation that shows the distribution of pitches (or fundamental frequencies) present in an audio signal over a specific duration.

Serrà, et al, 2011 have utilized interval histograms constructed from sung performances in Carnatic and Hindustani music show that Carnatic music tends toward just intonation, while Hindustani music exhibits more equal-tempered influences. Koduri et al, 2014, used performance pitch histograms and context-based svara distributions to compactly represent and analyze raga intonation.

Liu et al, 2022, applies similar techniques for analysis of Chinese Folk Music and show that pitch histograms reveal a distinctive bell-shaped distribution in traditional Chinese anhemitonic pentatonic folk songs, reflecting the cultural characteristics.

Scherbaum, et al, 2022, demonstrated how pitch histograms can efficiently summarize large dataset of traditional Georgian singing recordings, facilitate comparative analysis across performances, ensembles, and regions, and enable insights into traditional Georgian tuning systems.

**4.3 Pitch histogram computation and samples from the Cairo Congress recordings**

In this section we discuss the potential use of pitch histograms to perform a comparative study of regional variations in Arab music. The paper does not aim to provide a complete study but only a demonstration. Such a study is left to the expert scholars of Arab music theory.

In a scenario where multiple recordings are available for a given maqam, it is preferable to obtain a single (aggregate or mean) histogram representing the complete set of recordings and build analysis based on such a representation. For the computation of an aggregate histogram, a form of alignment is required in the pitch space since the instrument tuning may vary across recordings. While automatic alignment methods (such as that presented in Bozkurt, 2008) have been shown to be effective in most of the cases, errors still exist (typically exhibiting errors of a perfect fourth or fifth, similar to those found in automatic key detection) and may distort the mean histogram obtained if the number of recordings in the set are low. A more guaranteed way of aligning histograms is the use of tonic labels: representing/converting the histogram as/to an interval histogram (with respect to the tonic frequency). Automatic tonic identification methods have been proposed and successfully applied for Indian music (Gulati et al, 2014, Aiswarya & Rajan, 2024). Yet the algorithms are culture specific and are not free of errors either.

A group of recordings (64 tracks) of the Cairo Congress dataset has been manually labeled specifying the start and end time of a prominent tonic played in the recording. From this segmentation information a tonic frequency is estimated for the recording via i) gathering all pitch estimates from different algorithms for the segment, ii) sorting all pitch values, iii) leaving out the first and last thirds of pitch values keeping only the mid third and iv) computing the mean of these pitch values.

In Figure 2, we present example pitch histograms for one of the tracks available in the dataset. Rather than using absolute frequency values, we represent pitches in cents with respect to a reference frequency. The top subplot depicts histograms obtained using pitch series estimated with three different extractors and a mean histogram computed from these three histograms. A reference frequency of 55 Hz has been used in the computation of these histograms. The mid-subplot provides an octave warped version of the histograms. The bottom subplot is the octave warped mean histogram, this time represented relative to the tonic frequency which was estimated by the procedure defined in the paragraph above.

The last subplot in Figure 2 is particularly useful for tuning studies facilitating visual inspection of intervals on a grid of uniform steps (solid lines placed each 100 cents and dashed green lines each 20 cent steps). For this maqam and recording, the second and third degrees of the scale is representative of a very commonly used/observed minor third narrower than an equal tempered minor third and a second degree which is located close to the midpoint between the tonic and the third (which has been subject to many theoretical debates and often stated to vary based on the direction of the melodic movement). The histogram also provides information about which pitches/intervals are more frequently played, providing a clue for the "emphasized"/important pitches of the scale. The mid-subplot also includes peak locations and intervals automatically computed from the histograms, all in units of cents.

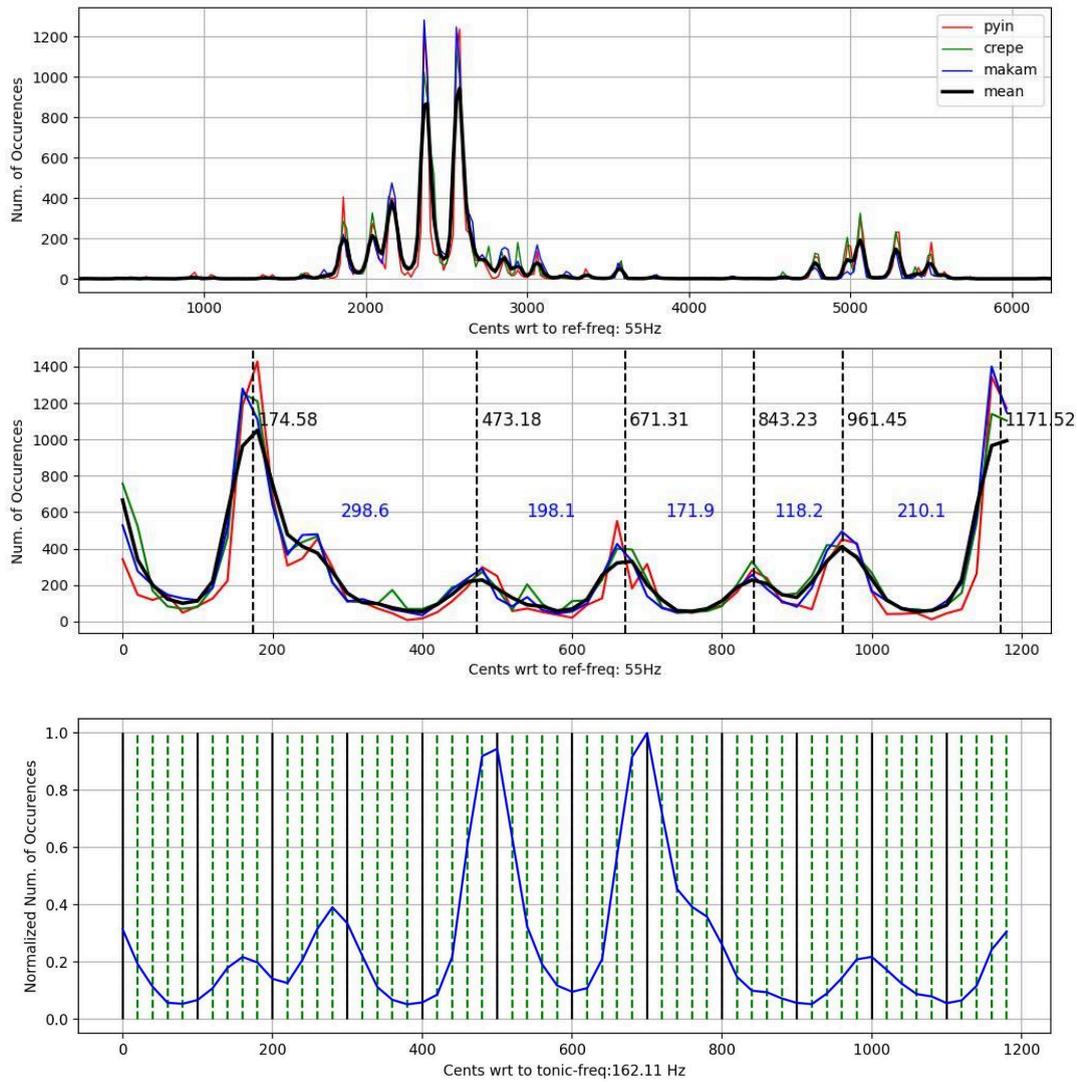

**Figure 2:** Pitch histogram plots for CD 08 - track 8, maqam: bayati, region: Egypt. top-subplot) histograms obtained using pitch series estimated with three different extractors and a mean of those histograms, mid-subplot) octave warped versions of the top-subplots, bottom-subplot) mean histogram represented with respect to the tonic frequency

The top two subplots in Figure 2 are created for all the recordings of the dataset and the bottom subplot is created for all the recordings with a tonic label. These plots are shared within the dataset (as jpeg files) to visual inspection without a dedicated analysis tool.

It is a well known nature of the maqam music culture that intervals vary due to musicians' personal choices, physical settings of the instrument (fretless/fretted). In addition to histograms of individual recordings, it is also useful to compute an aggregate histogram from a set of recordings in a specific maqam. The aggregation could be performed in various ways, the simplest being simple averaging of tonic aligned octave warped histograms. In Figure 3 and Figure 4, we present aggregate histograms for maqam rast and maqam husayni, grouping recordings with respect to their region. Vocal-only recordings are excluded in such mean histogram computations due to the risk of a potential pitch drift within the recording.

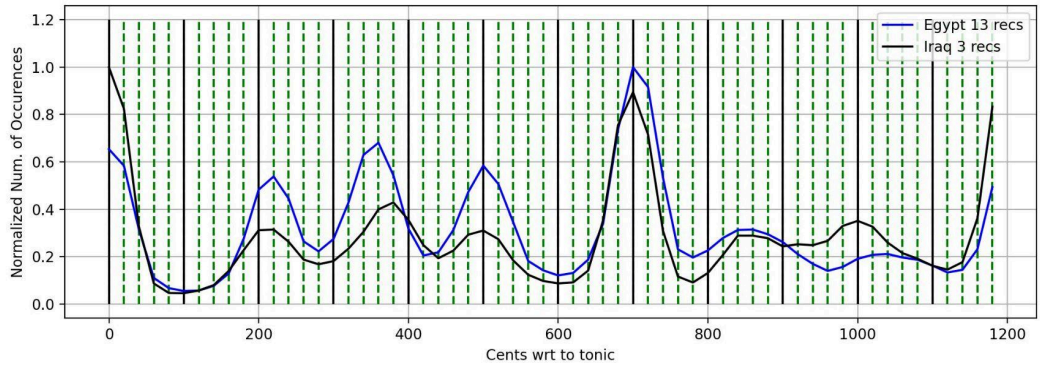

**Figure 3:** Mean histograms for maqam rast. The series are obtained by averaging tonic aligned histograms of 13 recordings (region: Egypt) and 3 recordings (Iraq).

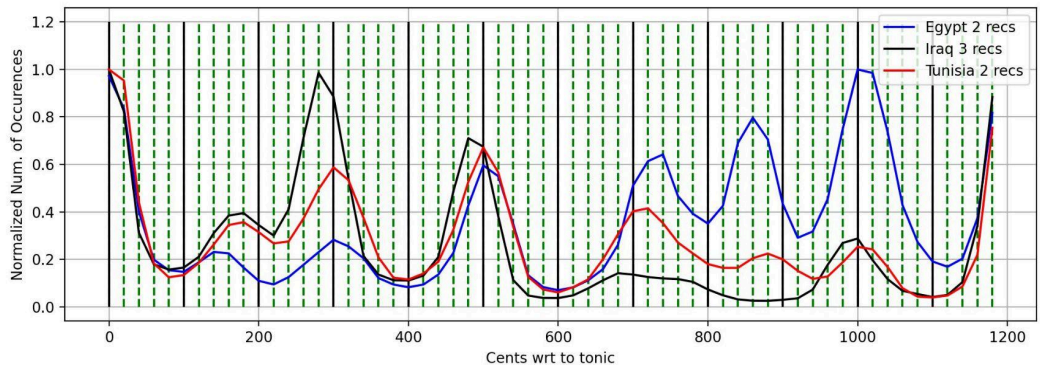

**Figure 4:** Mean histograms for maqam husayni. The series are obtained by averaging tonic aligned histograms of 2 recordings (region: Egypt), 3 recordings (Iraq) and 2 recordings (Tunisia).

Visually inspecting the figures 3 and 4, third degree of maqam Rast and the second degree of maqam Husayni favors that quarter-tone may indeed be a regional choice (Egypt) not applied in recordings from Iraq or Tunisia. We will leave further interpretation to expert scholars of Arab music.

## 5. Conclusion

This paper has introduced an open research dataset, ORD-CC32, centered on the 1932 Cairo Congress recordings, a pivotal and underutilized resource for Arab music studies. The dataset includes metadata, melodic and rhythmic mode annotations, tonic labels, and acoustic features, providing structured and machine-readable material. By transforming this historically and culturally significant collection into an accessible corpus, this work bridges a long-standing gap in computational approaches to Arab music traditions.

The value of this dataset lies in its rich cross-regional coverage and historical depth, enabling a wide range of computational inquiries. Researchers can leverage the dataset for tasks such as tuning and temperament analysis, melodic pattern characterization, form classification, rhythmic analysis, and modal structure exploration. These investigations are essential not only for algorithmic development and music information retrieval (MIR) applications (such as Arabic Music Genre Identification, Ahmed et al, 2024) but also for advancing theoretical and historical understanding of Arab music systems.

A notable use-case enabled by this dataset is the comparative analysis of tuning practices across different regions and historical periods. As demonstrated in section 4.3, the dataset provides the means to compute tonic-aligned pitch histograms and evaluate microtonal and modal variations in a data-driven manner. This supports studies akin to that of Kroher et al. (2018), which examined the tuning differences between flamenco and Arab-Andalusian vocal music. ORD-CC32 opens similar possibilities for comparing tuning norms across North Africa, Levantine, Iraqi, and Egyptian traditions as represented in the congress recordings.

Beyond analytical research, this dataset holds potential for interactive applications in education and cultural dissemination (such as immersive VR visualizations of music data, Ganguli,et al 2020, Guedes, 2023). The structured annotations and pitch features can be utilized to develop interactive music exploration platforms, pedagogical tools for maqam instruction, and digital exhibits for cultural institutions. Such tools can enhance both scholarly engagement and public appreciation of Arab music heritage, ensuring its transmission to future generations through accessible and technology-supported formats.

## Acknowledgement

This work has been funded by the Office of Research of Zayed University, United Arab Emirates, as part of Start-up Grants (Project no: 23035). All the maqam, rhythmic mode, musical form tags are added (primary resource: book of the congress publication Lambert, J, Cordereix, P. (Ed.). (2015)) to musicBrainz by Arthur Diniz De Souza, the undergraduate student research assistant of the project. Tonic segmentation has been performed by the author and Arthur Diniz De Souza.